\documentclass[11pt]{article}
\usepackage{fullpage}
\usepackage{amsmath}
\usepackage{amssymb}
\usepackage{amsfonts}
\usepackage[all,poly]{xy}
\usepackage[dvips]{color}

\newcounter{algoctr}

\newif\ifnotesw\noteswtrue
   {\ifnotesw\marginpar[\hfill\(\top\)]{\(\top\)}\fi}%
   {\ifnotesw\marginpar[\hfill\(\bot\)]{\(\bot\)}\fi}

\newcommand{\mnote}[1]%
    {\ifnotesw\marginpar%
        [{\scriptsize\begin{minipage}[t]{\marginparwidth}
        \raggedleft#1%
                        \end{minipage}}]%
        {\scriptsize\begin{minipage}[t]{\marginparwidth}
        \raggedright#1%
                        \end{minipage}}%
    \fi}

\newcommand{\ignore}[1]{}

\newsavebox{\given}
\savebox{\given}[1em]{\rule[-1.5ex]{.2mm}{4ex}}

\newcommand{\bnum}{\begin{equation}}
\newcommand{\enum}{\end{equation}}

\newtheorem{theorem}{Theorem}
\newtheorem{corollary}[theorem]{Corollary}
\newtheorem{lemma}[theorem]{Lemma}
\newtheorem{proposition}[theorem]{Proposition}

\newtheorem{fact}[theorem]{Fact}

\newtheorem{definition}{Definition}

\newcommand{\blackslug}{\rule{7pt}{7pt}}

\newcommand{\proof}{{\bf Proof: \enspace}}

\newcommand{\iverson}[1]{\lbrack\!\lbrack #1 \rbrack\!\rbrack}

\newcommand{\real}{\ifmmode {\rm R} \else ${\rm R}$ \fi}
\newcommand{\nat}{\ifmmode {\rm N} \else ${\rm N}$  \fi}
\newcommand{\tot}{\ifmmode {\cal T} \else ${\cal T}$ \fi}
\newcommand{\sigstar}{\ifmmode \Sigma^{\ast} \else $\Sigma^{\ast}$ \fi}

\newcommand{\inn}{\ifmmode \in \else $\in$ \fi}
\renewcommand{\phi}{\ifmmode \varphi \else $\varphi$ \fi}
\renewcommand{\le}{\ifmmode \leq \else $\leq$ \fi}
\renewcommand{\ge}{\ifmmode \geq \else $\geq$ \fi}
\renewcommand{\ne}{\ifmmode \neq \else $\neq$ \fi}
\newcommand{\lt}{\ifmmode < \else $<$ \fi}
\newcommand{\gt}{\ifmmode > \else $>$ \fi}
\newcommand{\eq}{\ifmmode = \else $=$ \fi}
\newcommand{\half}{\ifmmode \frac{1}{2} \else $\frac{1}{2}$ \fi}
\newcommand{\oneovern}{\ifmmode \frac{1}{n} \else $\frac{1}{n}$ \fi}

\newcommand{\ra}{\ifmmode \rightarrow \else $\rightarrow$ \fi}
\newcommand{\qed}{\hfill{\setlength{\fboxsep}{0pt}
\framebox[7pt]{\rule{0pt}{7pt}}}}

\renewcommand{\notin}{\ifmmode \not\in \else $\not\in$ \fi}
\newlength{\thislabel}
\newcommand{\labsize}[1]{\settowidth{\thislabel}{#1}}

\newcommand{\prf}{\par\noindent{\sl Proof } \hspace{.01 in}}

\newcommand{\lip}{\langle}
\newcommand{\rip}{\rangle}

\def\Complex{\mathbb C}
\def\Int{\mathbb Z}

\newcommand{\bra}[1]{\lip #1 |}
\newcommand{\ket}[1]{| #1 \rip}
\newcommand{\braket}[2]{\lip #1 | #2 \rip}

\newcommand{\dt}{\mathsf{d}t}

\newcommand{\crc}[1]{\lip #1 \rip}

\definecolor{purple}{rgb}{0.5,0,0.5}

\title{
Universal Mixing of Quantum Walk on Graphs\footnote{Supported in part by NSF grants DMR-0121146 and DMS-0353050.}
} 
\author{
{William Carlson}\\{\em Kansas State University}
\and {Allison Ford}\\{\em Mary Baldwin College}
\and {Elizabeth Harris}\\{\em SUNY Potsdam} 
\and {Julian Rosen}\\{\em University of Oklahoma}
\and {Christino Tamon}\\{\em Clarkson University}
\and {Kathleen Wrobel}\\{\em SUNY Potsdam}
}
\date{\today}

\begin{document}
\bibliographystyle{plain}
\maketitle

\begin{abstract}
We study the set of probability distributions visited by a continuous-time quantum walk on graphs. 
An edge-weighted graph $G$ is {\em universal mixing} if the instantaneous or average probability 
distribution of the quantum walk on $G$ ranges over all probability distributions on the vertices 
as the weights are varied over non-negative reals. The graph is {\em uniform} mixing if it visits 
the uniform distribution.
Our results include the following:
\begin{itemize}
\item All weighted complete multipartite graphs are instantaneous universal mixing. \\
	This is in contrast to the fact that no {\em unweighted} complete multipartite graphs 
	are uniform mixing (except for the four-cycle $K_{2,2}$).
\item For all $n \ge 1$, the weighted claw $K_{1,n}$ is a minimally connected instantaneous universal mixing graph. 
	In fact, as a corollary, the unweighted $K_{1,n}$ is instantaneous uniform mixing. This adds a new family
	of uniform mixing graphs to a list that so far contains only the hypercubes.
\item Any weighted graph is average almost-uniform mixing unless its spectral type is sublinear in the 
	size of the graph. This provides a nearly tight characterization for average uniform mixing on circulant graphs.
\item No weighted graphs are average universal mixing. This shows that weights do not help to achieve
	average universal mixing, unlike the instantaneous case.
\end{itemize}
Our proofs exploit the spectra of the underlying weighted graphs and path collapsing arguments.
\end{abstract}

\section{Introduction}

The theory of random walks on graphs is an important topic in mathematics, physics, and computer science 
\cite{spitzer, bollobas, feynman}. In recent years, a generalization of the classical random walks -- called
quantum walks -- has gained considerable interest in the quantum information and computation research areas
due to its potential applications \cite{aakv01}. 
In particular, the study of continuous-time quantum walks on graphs has shown promising applications in
the algorithmic and implementation aspects.
As an alternate algorithmic technique to the Quantum Fourier Transform and the Amplitude Amplification techniques, 
Childs et al. \cite{ccdfgs03} demonstrated the power of continuous-time quantum walk algorithm for solving a specific 
blackbox graph search problem. As an alternate model for quantum computation, continuous-time quantum walks
provide simple, yet ubiquitous in nature, promising physical realizations for quantum computers \cite{drkb02}.
To this end, analyses of decoherence in the quantum walk models have been carried out in several works
\cite{kendon2, kt03, ar05, fst06}.

In this paper, we study the set of probability distributions generated by continuous-time quantum walks on 
edge-weighted graphs. Previous works had studied the question of whether a quantum walk on certain graphs visits
the uniform distribution on the vertices of the graph \cite{mr02,abtw03,gw03}. 
Here, we consider graphs which visit {\em all} probability distributions on the vertex set of the graph. 
We call such graphs having the {\em universal mixing} property, whereas graphs that hit the uniform distribution 
have the {\em uniform mixing} property. We consider both the {\em instantaneous} and {\em average} distributions 
for such quantum walks. It is necessary to allow symmetric edge-weights on our graphs, since no unweighted graphs 
are universal mixing (although some, like the hypercubes, are uniform mixing \cite{mr02}).

Our study of universal mixing via quantum walks is motivated by recent works in random walks on graphs.
In \cite{kr04}, Kindler and Romik provided a characterization of the set of distributions computable by random
walks on finite state generators (directed graphs with outputs). In another set of works, Boyd, Diaconis, Sun, and Xiao 
\cite{bdx04, bdsx06} studied the problem of finding the set of edge weights on a fixed given graph so as to obtain 
the fastest mixing time for the random walk. In the context of these works, the main problem that we study is
as follows: given a fixed family of graphs, as we vary the edge weights on these graphs, will the quantum walk 
visit all probability distributions on the vertices? Stated differently, we are looking for a set of edge weights 
that allows the quantum walk to hit any specified probability distribution. Our main goal in this work is to discover 
and characterize graphs which allow such {\em universal} mixing property, as well as the more restricted uniform
mixing property.

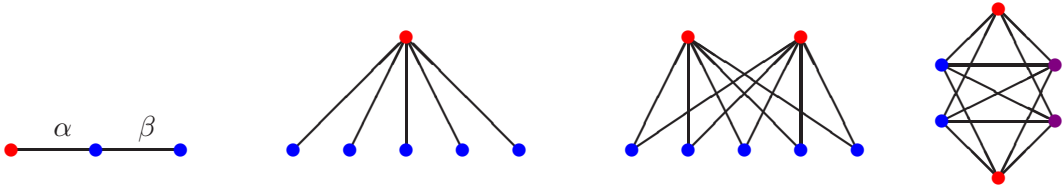
\begin{figure}[t]
\begin{center}
\setlength{\unitlength}{0.75cm}
\begin{picture}(24,4)
\thicklines

\put(1,1){\line(1,0){1.5}}
\put(2.5,1){\line(1,0){1.5}}
\put(1.75,1.25){\text{$\alpha$}}
\put(3.25,1.25){\text{$\beta$}}
\put(1,1){\color{red}{\circle*{0.25}}}
\put(2.5,1){\color{blue}{\circle*{0.25}}}
\put(4,1){\color{blue}{\circle*{0.25}}}

\put(6,1){\line(1,1){2}}
\put(7,1){\line(1,2){1}}
\put(8,1){\line(0,1){2}}
\put(9,1){\line(-1,2){1}}
\put(10,1){\line(-1,1){2}}
\put(6,1){\color{blue}{\circle*{0.25}}}
\put(7,1){\color{blue}{\circle*{0.25}}}
\put(8,1){\color{blue}{\circle*{0.25}}}
\put(8,3){\color{red}{\circle*{0.25}}}
\put(9,1){\color{blue}{\circle*{0.25}}}
\put(10,1){\color{blue}{\circle*{0.25}}}

\put(12,1){\line(1,2){1}}
\put(13,1){\line(0,1){2}}
\put(14,1){\line(-1,2){1}}
\put(15,1){\line(-1,1){2}}
\put(16,1){\line(-3,2){3}}
\put(12,1){\line(3,2){3}}
\put(13,1){\line(1,1){2}}
\put(14,1){\line(1,2){1}}
\put(15,1){\line(0,1){2}}
\put(16,1){\line(-1,2){1}}

\put(12,1){\color{blue}{\circle*{0.25}}}
\put(13,1){\color{blue}{\circle*{0.25}}}
\put(13,3){\color{red}{\circle*{0.25}}}
\put(14,1){\color{blue}{\circle*{0.25}}}
\put(15,1){\color{blue}{\circle*{0.25}}}
\put(15,3){\color{red}{\circle*{0.25}}}
\put(16,1){\color{blue}{\circle*{0.25}}}

\put(17.5,1.5){\line(1,0){2}}
\put(17.5,2.5){\line(1,0){2}}
\put(17.5,1.5){\line(2,1){2}}
\put(17.5,2.5){\line(2,-1){2}}
\put(17.5,1.5){\line(1,-1){1}}
\put(17.5,2.5){\line(1,-2){1}}
\put(19.5,1.5){\line(-1,-1){1}}
\put(19.5,2.5){\line(-1,-2){1}}
\put(17.5,1.5){\line(1,2){1}}
\put(17.5,2.5){\line(1,1){1}}
\put(19.5,1.5){\line(-1,2){1}}
\put(19.5,2.5){\line(-1,1){1}}

\put(17.5,1.5){\color{blue}{\circle*{0.25}}}
\put(17.5,2.5){\color{blue}{\circle*{0.25}}}
\put(18.5,0.5){\color{red}{\circle*{0.25}}}
\put(18.5,3.5){\color{red}{\circle*{0.25}}}
\put(19.5,1.5){\color{purple}{\circle*{0.25}}}
\put(19.5,2.5){\color{purple}{\circle*{0.25}}}

\end{picture}
\caption{Examples of edge-weighted graphs that are instantaneous universal mixing.
From left to right:
(a) path $P_{3}$; (b) claw $K_{1,5}$; (c) bipartite double-claw $K_{2,5}$;
(d) $4$-partite $K_{2,2,2,2}$.
}
\label{figure:universal}
\vspace{.1in}
\hrule
\end{center}
\end{figure}

First, we prove that complete multipartite graphs are instantaneous universal mixing. These are classes of graphs 
whose vertices are partitioned into disjoint sets, where all edges are present except for edges connecting vertices 
from the same partition.
In contrast, it is known that none of the unweighted complete multipartite graphs are uniform mixing, 
except for the four-cycle $K_{2,2}$ (see \cite{abtw03}). To show our multipartite theorem, we prove that the 
weighted three-vertex path $P_{3}$ and the claw (star) graph $K_{1,n}$ are both instantaneous universal mixing
(see Figure (\ref{figure:universal}) for examples of both graphs).
Our proofs employ a generalization of the {\em path collapsing} technique used in \cite{ccdfgs03}, 
adapted for weighted graphs. In \cite{ccdfgs03}, a path collapsing argument was used to show a fast hitting time 
of a continuous-time quantum walk on {\em glued tree} graphs; whereas, in this paper we use a generalization of 
the argument to show universal mixing on multipartite graphs.

In fact, the claw is a minimally connected graph that is universal mixing, since it forms a tree on the set of vertices. 
This shows that any graph with a claw subgraph is also instantaneous universal mixing. As a corollary, we observe that 
the {\em unweighted} claws are instantaneous uniform mixing. This adds a new family of uniform mixing graphs to a list 
that so far contains only the hypercubes \cite{mr02}.

Next, we consider a {\em closure} result on graphs with instantaneous uniform mixing. More specifically,
the Cartesian product $G \oplus H$ of two uniform mixing graphs $G$ and $H$ is also uniform mixing provided 
the two graphs share a common mixing time.
This is the fundamental property used to show that the hypercubes $Q_{n}$ are uniform mixing, since they are
the $n$-fold Cartesian product of the complete $2$-vertex graph $K_{2}$ with itself \cite{mr02}. 
We obtain several other classes of graphs with uniform mixing by combining the hypercubes $Q_{n}$ and the claws
$K_{1,n}$, for $n \ge 1$, the complete three-vertex and four vertex graphs ($K_{3}$ and $K_{4}$), using the 
Cartesian product operator. Since the three- and four-vertex cycles are equivalent to $K_{3}$ and $Q_{2}$, 
respectively, they are also uniform mixing. The status of the $n$-cycles $C_{n}$ is still open though; 
but we show that $C_{5}$ is {\em not} uniform mixing.

Finally, we prove that no weighted graphs are average universal mixing. Intuitively, this is because the
quantum walk never forgets its start vertex; or, more formally, the average probability weight of the 
start vertex is bounded away from zero. In the case of uniform mixing, we observe that a necessary condition
for a weighted graph to be an average uniform mixing is for its spectral type (the number of distinct
eigenvalues) to be linear in the size of the graph. This provides a nearly tight characterization for 
circulant graphs since these graphs are average almost-uniform mixing if their eigenvalues have bounded
multiplicities \cite{lrsstw06}. 

In this paper, our focus is on continuous-time quantum walks. For a more complete exposition on quantum walks, 
the interested reader is referred to the excellent surveys by Kendon and Kempe \cite{kendon1, kendon2, kempe}.

\section{Preliminaries}

{\em Notation: } For a logical statement $S$, the Iversonian $\iverson{S}$ (introduced in \cite{gkp}) 
denotes the characteristic function of $S$ which evaluates to $1$ if $S$ is true, and to $0$ if it is false.

We consider graphs $G=(V,E)$ that are simple (no self-loops) and undirected, with edge weights. 
The edge weights are given by a non-negative real-valued function $\alpha: E \rightarrow \mathbb{R}^{+} \cup \{0\}$
that is symmetric, i.e., $\alpha_{j,k} = \alpha_{k,j}$, for all $j,k \in V$.
Let $A_{G}$ be the adjacency matrix of $G$, where $A_{G}[j,k] = \alpha_{j,k} \iverson{(j,k) \in E}$. 
The set of eigenvalues of $A_{G}$ is denoted $Sp(G)$, and the (algebraic) multiplicity of an eigenvalue 
$\lambda$ is denoted $m(\lambda)$.
The spectral {\em type} $\tau(G)$ of a graph $G$ is the number of distinct eigenvalues of the adjacency
matrix $A_{G}$ of $G$. The maximum (algebraic) multiplicity of any eigenvalue of graph $G$ is denoted $\mu(G)$.

Some of the families of graphs studied here include paths $P_{n}$, cycles $C_{n}$, hypercubes $Q_{n}$,
complete graphs $K_{n}$, complete multipartite graphs $K^{(k)}_{n}$, and circulant graphs. 
A complete multipartite graph $K^{(k)}_{n}$ is the graph complement of $k$ disjoint complete graphs $K_{n}$.
A graph is a {\em circulant} graph if its adjacency matrix is a circulant matrix. 
The Cartesian product of two graphs $G$ and $H$, denoted $G \oplus H$, is the graph defined
on the vertex set $G \times H$, where $(g_{1},h_{1})$ is adjacent to $(g_{2},h_{2})$ if 
$g_{1} = g_{2}$ and $(h_{1},h_{2}) \in E(H)$; or $(g_{1},g_{2}) \in E(G)$ and $h_{1} = h_{2}$ 
(see page 617, \cite{lovasz}).
Further background on graphs and their spectral properties are given in \cite{bollobas, biggs}.

\ignore{
A circulant matrix $A$ is specified by its first row, say $(a_{0}, a_{1}, \ldots, a_{n-1})$, and 
is defined as $A_{j,k} = a_{k-j \pmod{n}}$, where $j,k \in \Int_{n}$. Here $\Int_{n}$ denotes the group
of integers $\{0,\ldots,n-1\}$ under addition modulo $n$.
Note that $a_{0} = 0$, since our graphs are simple, and $a_{j} = a_{n-j}$, since our graphs are undirected.
Connectivity is guaranteed if the greatest common divisor of $n$ and all indices $k$, for which $a_{k}=1$, is one.
Alternatively, a circulant graph $G = (V,E)$ can be specified by a subset $S \subseteq \Int_{n}$, where $(j,k) \in E$
if $k-j \in S$. In this case, we write $G = \crc{S}$. We will assume that $S$ is closed under taking inverses,
namely, if $d \in S$, then $-d \in S$.
}

\ignore{
It is known that circulant graphs $G$ are diagonalizable by the Fourier matrix $F$ defined as
$F_{j,k} = n^{-1/2} \omega_{n}^{jk}$, where $\omega_{n} = \exp(2\pi i/n)$. In fact, the eigenvalues of $A_{G}$ are
\begin{equation} \label{eqn:eigenvalue}
\lambda_{j} = \sum_{k=1}^{n-1} a_{k}\omega^{jk} 
	= \sum_{k=1}^{\lfloor (n-1)/2 \rfloor} 2\cos\left(\frac{2\pi jk}{n}\right) \
		+ \ \iverson{n \mbox{ even}} \ a_{n/2} \ (-1)^{j}.
\end{equation}
}

A {\em continuous-time quantum walk} on a graph $G=(V,E)$ is defined using the Schr\"{o}dinger equation 
with the real symmetric matrix $A_{G}$ as the Hamiltonian (see \cite{ccdfgs03}). 
If $\ket{\psi(t)} \in \Complex^{|V|}$ is a time-dependent amplitude vector on the vertices of $G$, 
then the evolution of the quantum walk is given by
\begin{equation}
\ket{\psi(t)} = e^{-it A_{G}} \ket{\psi(0)},	
\end{equation}
where $i = \sqrt{-1}$ and $\ket{\psi(0)}$ is the initial amplitude vector. We usually assume that $\ket{\psi(0)}$ 
is a unit vector, with $\braket{x}{\psi(0)} = \iverson{x = \mbox{\sc start}}$, for some vertex $\mbox{\sc start}$.
The amplitude of the quantum walk of vertex $j$ at time $t$ is given by $\psi_{j}(t) = \braket{j}{\psi(t)}$.
The {\em instantaneous} probability of vertex $j$ at time $t$ is $p_{j}(t) = |\psi_{j}(t)|^{2}$. 
The {\em average} probability of vertex $j$ is defined as 
\begin{equation}
\overline{p}_{j} = \lim_{T \rightarrow \infty} \frac{1}{T} \int_{0}^{T} p_{j}(t) \ \dt.
\end{equation}
The average probability distribution of the quantum walk will be denoted $\overline{P}$. 
This notion of average distribution (defined in \cite{aakv01} for discrete-time quantum walks)
is similar to the notion of a stationary distribution in classical random walks \cite{aldous-fill}. 

\begin{definition} (Universal and Uniform Mixing) \\
Let $G=(V,E)$ be a simple, undirected, and connected graph that is edge-weighted. 
Then, $G$ has the instantaneous (or average) {\em universal mixing} property if for any probability 
distribution $Q$ over the vertex set $V$ and for any start vertex $x$, there is a set of non-negative real 
weights on $E$, so that the continuous-time quantum walk on the weighted $G$, starting from $x$, has an 
instantaneous probability distribution at time $t$ (or average distribution) that equals $Q$.

If the above condition holds for $Q$ being the uniform distribution on $V$, we say $G$ has the 
instantaneous (or average) {\em uniform mixing} property. The mixing is {\em almost-uniform} if 
the instantaneous (or average) probability of each vertex is at most $O(1/|V|)$.
\end{definition}

\par\noindent{\bf Example:}
A quantum walk on the connected $2$-vertex graph $K_{2}$ is given by
\begin{equation}
\exp\left(-it\begin{bmatrix} 0 & 1 \\ 1 & 0 \end{bmatrix}\right)\begin{bmatrix} 1 \\ 0 \end{bmatrix}
= \begin{bmatrix} \cos(t) \\ -i\sin(t) \end{bmatrix}.
\end{equation}
Thus, the instantaneous probability distribution of the quantum walk is $p(t) = [\cos^{2}(t) \sin^{2}(t)]^{T}$.
This shows that the quantum walk on $K_{2}$ can generate {\em any} probability distribution on the two vertices.
Unfortunately, this case does not generalize to arbitrarily many vertices. It was shown in \cite{abtw03} that 
the instantaneous probability distribution quantum walk on the complete graph $K_{n}$ {\em never} visits the 
uniform distribution on $n$ vertices, for any $n > 4$. 
A main question considered in this work is: will the quantum walk visit the uniform distribution if edge weights 
are allowed? In fact, as we vary the edge weights on $K_{n}$, will the quantum walk visit {\em all} probability 
distributions on $n$ elements (as is the case with the unweighted $K_{2}$)? 
We answer both questions in this paper; moreover, we will exhibit a family of minimally connected graphs with
such universal property. Note that in a classical random walk, the interference phenomenon commonly observed in
a quantum walk does not exist; thus, it is impossible for vertices reachable from the start vertex to have a zero
probability.


\section{Instantaneous Universal Mixing}

In this section we prove that all weighted complete multipartite graphs are instantaneous universal mixing.
First, we prove some results about the weighted $3$-path $P_{3}$ and claw $K_{1,n}$.

\begin{lemma} \label{lemma:P3}
The weighted $P_{3}$ has instantaneous universal mixing.
\end{lemma}
\proof
Without loss of generality, we assume that the weights on $P_{3}$ are $1$ and $\alpha$;
since we can always scale the first weight to unity. Let $A$ be the adjacency matrix of $G$.


\begin{figure}[h]
\begin{center}
\setlength{\unitlength}{0.75cm}
\begin{picture}(15,1.5)
\thicklines

\put(0,0){\text{$P_{3}:$}}
\put(2,0.5){\line(1,0){1.5}}
\put(3.5,0.5){\line(1,0){1.5}}
\put(2.55,0){\text{$1$}}
\put(4.05,0){\text{$\alpha$}}
\put(2,0.5){\color{blue}{\circle*{0.25}}}
\put(3.5,0.5){\color{blue}{\circle*{0.25}}}
\put(5,0.5){\color{blue}{\circle*{0.25}}}
\put(1.25,1){\text{\footnotesize\sc left}}
\put(2.75,1){\text{\footnotesize\sc middle}}
\put(4.55,1){\text{\footnotesize\sc right}}

\put(9,0.5){\text{$A = \begin{bmatrix}
	0 & 1 & 0 \\
	1 & 0 & \alpha \\
	0 & \alpha & 0
	\end{bmatrix}$}}
\end{picture}
\end{center}
\end{figure}

\ignore{
Thus, the adjacency matrix is:
\begin{equation}
A = \begin{bmatrix}
	0 & 1 & 0 \\
	1 & 0 & \alpha \\
	0 & \alpha & 0
	\end{bmatrix}
\end{equation}
}
\par\noindent
The eigenvalues of $A$ are $\lambda_{0} = 0$ and $\lambda_{\pm} = \pm \Delta$, where $\Delta = \sqrt{1 + \alpha^{2}}$,
with the following set of orthonormal eigenvectors:
\begin{equation}
\ket{v_{0}} = \frac{1}{\Delta}\begin{bmatrix} -\alpha \\ 0 \\ 1 \end{bmatrix},
\ \ \ \hspace{.2in} \ \ \
\ket{v_{\pm}} = \frac{1}{\sqrt{2\Delta^{2}}}\begin{bmatrix} 1 \\ \pm\Delta \\ \alpha \end{bmatrix},
\end{equation}
We have two cases to consider depending on the starting vertex of the quantum walk.

\paragraph{case A:}
The quantum walk starting at the left vertex is given by:
\begin{equation}
e^{-itA}\ket{\mbox{\small\sc left}}
	= \frac{-\alpha}{\Delta}\ket{v_{0}} + \frac{1}{\sqrt{2\Delta^{2}}}\sum_{\pm} e^{\mp it\Delta}\ket{v_{\pm}}
	= \frac{1}{\Delta^{2}}
		\begin{bmatrix}
		(\alpha^{2} + \cos(\Delta t)) \\
		-i\Delta\sin(\Delta t) \\
		\alpha(\cos(\Delta t)-1)
		\end{bmatrix}
\end{equation}
Thus, the instantaneous probability distribution at time $t$ is:
\begin{equation}
p_{\mbox{\scriptsize\sc left}}(t) 
	= (1 - 2\Gamma)^{2}, \ \ \
p_{\mbox{\scriptsize\sc middle}}(t) 	
	= 4\Gamma(1-\Gamma\Delta^{2}), \ \ \
p_{\mbox{\scriptsize\sc right}}(t) 	
	= \alpha^{2}(2\Gamma)^{2},
\end{equation}
where $\Gamma = \sin^{2}(\Delta t/2)/\Delta^{2}$. 
Combining the first and third expressions, we get
$\alpha = \sqrt{p_{\mbox{\scriptsize\sc right}}(t)}/(1 - \sqrt{p_{\mbox{\scriptsize\sc left}}(t)})$,
which shows that $(\alpha, t)$ can be selected to satisfy any probability distribution on the three vertices.

\paragraph{case B:}
The quantum walk starting at the middle vertex is given by:
\begin{equation}
e^{-itA}\ket{\mbox{\small\sc middle}} 
	= \frac{1}{\sqrt{2\Delta^{2}}}\sum_{\pm} (\pm\Delta) e^{\mp it\Delta}\ket{v_{\pm}}
	= \frac{1}{\Delta}
		\begin{bmatrix}
		-i\sin(\Delta t)) \\
		\Delta\cos(\Delta t) \\
		-i\alpha\sin(\Delta t)
		\end{bmatrix}
\end{equation}
Thus, the instantaneous probability distribution at time $t$ is:
\begin{equation}
p_{\mbox{\scriptsize\sc left}}(t) 	= \frac{\sin^{2}(\Delta t)}{\Delta^{2}}, \ \ \
p_{\mbox{\scriptsize\sc middle}}(t) 	= \cos^{2}(\Delta t), \ \ \
p_{\mbox{\scriptsize\sc right}}(t) 	= \alpha^{2}\frac{\sin^{2}(\Delta t)}{\Delta^{2}}.
\end{equation}
Thus $\alpha = \sqrt{p_{\mbox{\scriptsize\sc right}}}/\sqrt{p_{\mbox{\scriptsize\sc left}}}$, and hence,
we see that $(\alpha, t)$ can be chosen to satisfy any required probability triples.
\qed \\

\par\noindent
In the following, we show that the weighted claw (star) graph is instantaneous universal mixing,
for an arbitrary starting vertex. We will use Lemma \ref{lemma:P3} to prove this in combination with a
weighted version of the {\em path collapsing} argument (used in \cite{ccdfgs03}).

\begin{theorem} \label{theorem:claw}
The weighted $K_{1,n}$ has instantaneous universal mixing, for $n \ge 1$.
Moreover, the weighted complete graphs $K_{n}$ are also instantaneous universal mixing, for $n \ge 1$.
\end{theorem}
\prf
Let the edge weights on the claw be $\alpha_{1},\ldots,\alpha_{n}$, respectively. 
Then, the adjacency matrix is given by:
\begin{equation}
A = 	\begin{bmatrix}
	0 & \alpha_{1} & \alpha_{2} & \ldots & \alpha_{n} \\
	\alpha_{1} & 0 & 0 & \ldots & 0 \\
	\alpha_{2} & 0 & 0 & \ldots & 0 \\
	\vdots & \vdots & \vdots & \vdots & \vdots \\
	\alpha_{n} & 0 & 0 & \ldots & 0
	\end{bmatrix}
\end{equation}
The eigenvalues of $A$ are $\lambda_{\pm} = \pm\Delta$, where $\Delta = \sqrt{\sum_{k=1}^{n} \alpha_{k}^{2}}$,
and $\lambda_{0} = 0$. The eigenvalues $\lambda_{\pm}$ are simple, whereas $0$ has multiplicity $n-1$.
The eigenvectors are given by:
\begin{eqnarray}
\ket{v_{\pm}} & = & 
	\frac{1}{\sqrt{2}} \begin{bmatrix} 1 & \pm\alpha_{1}/\Delta & \ldots & \pm\alpha_{n}/\Delta \end{bmatrix}^{T} \\
\ket{v_{0}} & = & \begin{bmatrix} 0 & y_{1} & \ldots & y_{n} \end{bmatrix}^{T}, \ \ \ 
	\mbox{ where $\sum_{k=1}^{n} \alpha_{k} y_{k} = 0$}
\end{eqnarray}
Depending on whether the quantum walk starts at the center of the claw or not, we have two cases to analyze.

\paragraph{case A:}
The quantum walk starting at the center of the claw is given by:
\begin{equation}
\ket{\psi(t)} = e^{-itA} \sum_{\pm} \frac{1}{\sqrt{2}} \ket{v_{\pm}}
\end{equation}
which yields
$\braket{\mbox{\small\sc center}}{\psi(t)} = \cos(\Delta t)$, and
$\braket{k}{\psi(t)} = -i\alpha_{k}\sin(\Delta t)/\Delta$, for $k=1,\ldots,n$.
Thus, the instantaneous probabilities are given by:
\begin{equation}
p_{\mbox{\scriptsize\sc center}}(t) = \cos^{2}(\Delta t), \ \ \
p_{k}(t) = \frac{\alpha_{k}^{2}}{\Delta^{2}}\sin^{2}(\Delta t), \ \ \mbox{ where $k=1,\ldots,n$ }
\end{equation}
This shows that the above instantaneous probabilities ranges over all probability distributions on $n+1$ vertices
as $t$ and the $\alpha_{k}$'s range over $\mathbb{R}^{+} \cup \{0\}$.

\begin{figure}[t]
\begin{center}
\setlength{\unitlength}{0.75cm}
\begin{picture}(15,5)
\thicklines
\put(1,2){\line(2,1){2}}
\put(1,2){\line(1,0){2}}
\put(1,2){\line(2,-1){2}}
\put(1,2){\color{blue}{\circle*{0.25}}}
\put(3,1){\color{purple}{\circle*{0.25}}}
\put(3,3){\color{purple}{\circle*{0.25}}}
\put(3,2){\color{red}{\circle*{0.25}}}

\put(0,2.5){\text{$p_{\mbox{\tiny\sc center}}$}}
\put(3.5,3.05){\text{$p_{2}$}}
\put(3.5,2){\text{$p_{1}$}}
\put(3.5,0.95){\text{$p_{3}$}}

\put(5.5,2){\text{$\Longrightarrow$}}

\put(12,2){\color{blue}{\oval(1,3.25)}}

\put(8,2){\line(1,0){2}}
\put(10,2){\line(2,1){2}}
\put(10,2){\line(2,-1){2}}
\put(10,2){\color{blue}{\circle*{0.25}}}
\put(12,1){\color{purple}{\circle*{0.25}}}
\put(12,3){\color{purple}{\circle*{0.25}}}
\put(8,2){\color{red}{\circle*{0.25}}}

\put(8.15,1.45){\text{$\sqrt{p_{\mbox{\tiny\sc center}}}$}}
\put(10.5,3){\text{$\sqrt{p_{2}}$}}
\put(10.5,0.85){\text{$\sqrt{p_{3}}$}}

\put(7.5,2.35){\mbox{\small\sc left}}
\put(9.05,2.35){\mbox{\small\sc middle}}
\put(11.45,3.95){\mbox{\small\sc right}}

\end{picture}
\caption{Case B: the claw $K_{1,3}$ is universal mixing, when $\mbox{\sc start} \neq \mbox{\sc center}$;
a reduction to $P_{3}$. The start vertex is given in red and the target probabilities (left) are {\em shifted} 
onto the edges of the graph (right).
}
\label{figure:claw-to-path}
\vspace{.1in}
\hrule
\end{center}
\end{figure}
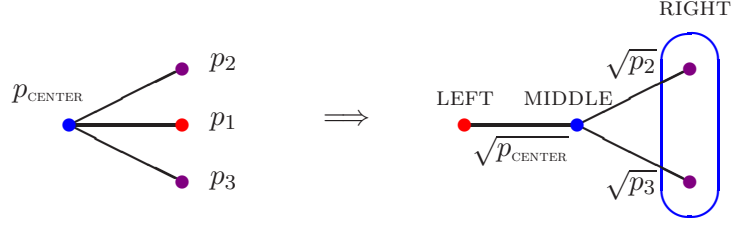

\paragraph{case B:}
We can assume without loss of generality that the quantum walk starts at vertex $k=1$.
But, this case is similar to the weighted $P_{3}$ case where vertex $1$ is $\mbox{\small\sc left}$, the center of the
claw is $\mbox{\small\sc middle}$, and the rest of the other vertices are viewed as $\mbox{\small\sc right}$;
see Figure \ref{figure:claw-to-path}.
A more formal argument for this reduction is as follows. 
Given the target probabilities $p_{1}$, $p_{\mbox{\scriptsize\sc center}}$, and $p_{2},\ldots,p_{n}$, we define the weights
on $K_{1,n}$ as follows: $w(1,\mbox{\scriptsize\sc center})=\sqrt{p_{\mbox{\scriptsize\sc center}}}$ and 
$w(\mbox{\small\sc center},k) = \alpha_{k}$, where $\alpha_{k} = \sqrt{p_{k}}$, for $k=2,\ldots,n$.
Along with the states $\ket{\mbox{\small\sc left}} = \ket{1}$ and $\ket{\mbox{\small\sc middle}} = 
\ket{\mbox{\small\sc center}}$, 
we define a new state:
\begin{equation}
\ket{\mbox{\small\sc right}} = \sum_{k=2}^{n} \frac{\alpha_{k}}{\widetilde{\Delta}} \ket{k}, \ \ \
\mbox{ where $\widetilde{\Delta} = \sqrt{\sum_{k=2}^{n} \alpha_{k}^{2}}$}. 
\end{equation}
Under this new reduced basis, the quantum walk
on $K_{1,n}$, starting at vertex $1$, is expressed using a {\em collapsed} Hamiltonian on $P_{3}$:
\begin{equation}
\ket{\Psi(t)} = 
	\exp\left(-it
	\begin{bmatrix}
	0 & 1 & 0 \\
	1 & 0 & \widetilde{\Delta} \\
	0 & \widetilde{\Delta} & 0
	\end{bmatrix}\right)
	\begin{bmatrix} 1 \\ 0 \\ 0 \end{bmatrix}
\end{equation}

Note that the amplitudes $\braket{k}{\Psi(t)}$ in the original $K_{1,n}$ is proportional to the amplitude 
$\braket{\mbox{\small\sc right}}{\Psi(t)}$, where the constant of proportionality is given by $\alpha_{k}$. 
Next, we find a mixing time $T$ on the weighted $P_{3}$ with the probabilities $p_{\mbox{\scriptsize\sc left}}(T) 
= p_{1}$, $p_{\mbox{\scriptsize\sc middle}}(T) = p_{\mbox{\scriptsize\sc center}}$, and 
$p_{\mbox{\scriptsize\sc right}}(T) = \sum_{k=2}^{n} p_{k}$.
At time $T$, the probability of vertex $k$ in $K_{1,n}$ is 
$p_{k}(T) = \alpha_{k}^{2}/\Delta^{2} \times p_{\mbox{\scriptsize\sc right}}(T)$,
which equals the target probability $p_{k}$, for all $k=2,\ldots,n$.
\qed \\

\par\noindent
For the next result, we generalize the previous theorem on $K_{1,n}$ to arbitrary complete multipartite graphs.

\begin{theorem} \label{theorem:bipartite}
All weighted complete bipartite graphs $K_{m,n}$ are instantaneous universal mixing, for all $m,n \ge 1$.
\end{theorem}
\prf
If $m=1$, $K_{m,n}$ which is universal mixing by Theorem \ref{theorem:claw}. Now, assume that $m > 1$.
Let $A = \{a_{0},a_{1},\ldots,a_{m}\}$ and $B = \{b_{1},\ldots,b_{n}\}$ be the two partitions of the 
bipartite graph $G = K_{m+1,n}$, with $|A|=m+1$ and $|B|=n$.
Without loss of generality, let the start vertex be $a_{0}$.
Viewing the start vertex as its own partition, we have a weighted $3$-path where $a_{0}$, $B$ and 
$C = A\setminus\{a_{0}\}$ form the {\em vertices} of $P_{3}$. 

\begin{figure}[t]
\begin{center}
\setlength{\unitlength}{0.75cm}
\begin{picture}(15,5)
\thicklines

\put(2,1){\line(1,0){2}}
\put(2,1){\line(2,1){2}}
\put(2,1){\line(1,1){2}}

\put(2,2){\line(2,1){2}}
\put(2,2){\line(1,0){2}}
\put(2,2){\line(2,-1){2}}
\put(2,2){\color{red}{\circle*{0.25}}}

\put(2,3){\line(1,0){2}}
\put(2,3){\line(2,-1){2}}
\put(2,3){\line(1,-1){2}}

\put(2,1){\color{purple}{\circle*{0.25}}}
\put(2,3){\color{purple}{\circle*{0.25}}}
\put(4,1){\color{blue}{\circle*{0.25}}}
\put(4,2){\color{blue}{\circle*{0.25}}}
\put(4,3){\color{blue}{\circle*{0.25}}}

\put(1.15,2){\text{$q_{0}$}}
\put(1.15,1){\text{$q_{2}$}}
\put(1.15,3){\text{$q_{1}$}}
\put(4.45,1){\text{$p_{3}$}}
\put(4.45,2){\text{$p_{2}$}}
\put(4.45,3){\text{$p_{1}$}}

\put(6.95,2){\text{$\Longrightarrow$}}

\put(12,2){\color{blue}{\oval(1,3.25)}}
\put(14,2){\color{blue}{\oval(1,3.25)}}

\put(14,1){\line(-1,0){2}}
\put(14,1){\line(-2,1){2}}
\put(14,1){\line(-1,1){2}}

\put(14,3){\line(-1,0){2}}
\put(14,3){\line(-2,-1){2}}
\put(14,3){\line(-1,-1){2}}

\put(10,2){\line(2,1){2}}
\put(10,2){\line(1,0){2}}
\put(10,2){\line(2,-1){2}}
\put(10,2){\color{red}{\circle*{0.25}}}

\put(12,1){\color{blue}{\circle*{0.25}}}
\put(12,2){\color{blue}{\circle*{0.25}}}
\put(12,3){\color{blue}{\circle*{0.25}}} 
\put(14,1){\color{purple}{\circle*{0.25}}}
\put(14,3){\color{purple}{\circle*{0.25}}}

\put(10.25,2.85){\text{$\sqrt{p_{1}}$}}
\put(10.75,2.15){\text{$\sqrt{p_{2}}$}}
\put(10.25,1){\text{$\sqrt{p_{3}}$}}

\put(14,3.15){\color{red}{\oval(1.25,1.25)[lb]}}
\put(14.25,2.5){\text{$\sqrt{q_{1}p_{k}}$}}

\put(14,0.85){\color{red}{\oval(1.25,1.25)[lt]}}
\put(14.25,1.25){\text{$\sqrt{q_{2}p_{k}}$}}

\put(8.95,2.35){\text{\small\sc left}}
\put(11.25,3.95){\text{\small\sc middle}}
\put(13.5,3.95){\text{\small\sc right}}

\end{picture}
\caption{The complete bipartite graph $K_{3,3}$ is universal mixing: by a reduction to $P_{3}$.
The start vertex is given in red. The target probabilities $q_{0},q_{1},q_{2}$ and $p_{1},p_{2},p_{3}$ (left)
are transferred onto the edge weights of the graph (right).
}
\label{figure:bipartite-to-path}
\vspace{.1in}
\hrule
\end{center}
\end{figure}
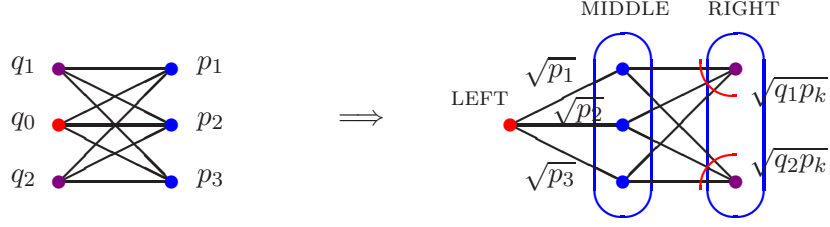

Let $p_{1},\ldots,p_{n}$ be the required probabilities on the vertices of $B$ 
and let $q_{1},\ldots,q_{m}$ be the required probabilities on the vertices of $C$.
Let $\alpha_{j} = \sqrt{p_{j}}$, for $j=1,\ldots,n$, and $\beta_{k} = \sqrt{q_{k}}$, for $k=1,\ldots,m$,
with $\Delta = \sqrt{\sum_{j=1}^{n} \alpha_{j}^{2}}$ and $\Gamma = \sqrt{\sum_{k=1}^{m} \beta_{k}^{2}}$.
Now, we define the following edge weights on $G$:
\begin{eqnarray}
w(a_{0},b_{j}) & = & \alpha_{j}, \ \ \ \mbox{ where $j=1,\ldots,n$ } \\
w(b_{j},a_{k}) & = & \alpha_{j}\beta_{k}, \ \ \ \mbox{ where $j=1,\ldots,n$ and $k=1,\ldots,m$ },
\end{eqnarray}
while the other weights are zero.
Consider the following quantum states
\begin{equation}
\ket{\mbox{\small\sc left}} = \ket{a_{0}}, 
\ \ \ 
\ket{\mbox{\small\sc middle}} = \frac{1}{\Delta}\sum_{j=1}^{n} \alpha_{j}\ket{a_{j}}, 
\ \ \
\ket{\mbox{\small\sc right}} = \frac{1}{\Gamma}\sum_{k=1}^{m} \beta_{k}\ket{b_{k}}.
\end{equation}
Under the basis states $\{\ket{\mbox{\sc left}},\ket{\mbox{\sc middle}},\ket{\mbox{\sc right}}\}$, 
we have the following {\em collapsed} Hamiltonian for a weighted $P_{3}$:
\begin{equation}
\mathbb{H} = \begin{bmatrix}
		0 & \Delta & 0 \\
		\Delta & 0 & \Delta\Gamma \\
		0 & \Delta\Gamma & 0
		\end{bmatrix}
\end{equation}
In the quantum walk $\ket{\Psi(t)} = \exp(-it\mathbb{H})\ket{a_{0}}$, note that
the amplitude $\braket{b_{j}}{\Psi(t)}$ in the original $K_{m,n}$ is proportional to
the amplitude $\braket{\mbox{\small\sc middle}}{\Psi(t)}$ by the constant $\alpha_{j}$, 
whereas the amplitude $\braket{a_{k}}{\Psi(t)}$ is proportional to the amplitude 
$\braket{\mbox{\small\sc right}}{\Psi(t)}$ by the constant $\beta_{k}$.

In the weighted $P_{3}$, we find a mixing time $T$ for which $p_{\mbox{\scriptsize\sc middle}}(T) = \sum_{j=1}^{n} p_{j}$
and $p_{\mbox{\scriptsize\sc right}}(T) = \sum_{k=1}^{m} q_{k}$. At this time $T$, the probability of vertex $b_{j}$ is
given by $\alpha_{j}^{2}/\Delta^{2} \times p_{\mbox{\scriptsize\sc middle}}(T) = p_{j}$, and the probability of vertex
$a_{k}$ is given by $\beta_{k}^{2}/\Gamma^{2} \times p_{\mbox{\scriptsize\sc right}}(T) = q_{k}$. 
This completes the claim.
\qed \\

\begin{theorem} \label{theorem:multipartite}
All weighted complete $k$-partite graphs are instantaneous universal mixing, for $k \ge 2$.
\end{theorem}
\prf
We prove the claim by induction on $k$. For $k=2$, we have a complete bipartite graph which is universal
mixing by Theorem \ref{theorem:bipartite}.
Assuming the claim is true for all $k < t$, any complete $t$-partite graph contains a complete $(t-1)$-partite 
subgraph (by disconnecting two arbitrary partitions), and thus, it is universal mixing. This proves the claim.
\qed \\

\par\noindent
The above theorem stands in contrast to the fact that (unweighted) complete multipartites, with the exception
of $K_{2,2}$, are not instantaneous uniform mixing (see \cite{abtw03}).


\section{Instantaneous Uniform Mixing}
The only unweighted graphs known to be uniform mixing are the hypercubes $Q_{n}$ \cite{mr02} and the two 
complete graphs, $K_{3}$ and $K_{4}$ \cite{abtw03}. To this small list, we add another family of graphs.

\begin{corollary}
The family of (unweighted) claw $K_{1,n}$ graphs is instantaneous uniform mixing.
\end{corollary}
\prf
Apply Theorem \ref{theorem:claw} with $\alpha_{k}=1$, $k=1,\ldots,n$, and $t = \cos^{-1}(1/\sqrt{n+1})/\sqrt{n}$.
\qed \\

\par\noindent
In what follows, we state a {\em closure} result for graphs that are uniform mixing.

\begin{fact} \label{fact:cartesian_product}
If $G,H$ are graphs with instantaneous uniform mixing, then so is $G \oplus H$, assuming that their
mixing times have a common intersection.
\end{fact}
\prf
Let $\{\langle \mu_{j},\ket{v_{j}}\rangle\}_{j}$ and $\{\langle \nu_{k},\ket{w_{k}}\rangle\}_{k}$ 
be the spectra of $G$ and $H$, respectively. 
The adjacency matrix of $G \oplus H$ is given by $I \otimes G + H \otimes I$, which 
is a sum of two commuting matrices. Hence, $\ket{v_{j}}\otimes\ket{w_{k}}$ are the eigenvectors of 
$G \oplus H$ with eigenvalues $\mu_{j} + \nu_{k}$, for all $j,k$.
Without loss of generality, assume that the start vertex is $\ket{0}_{G} \otimes \ket{0}_{H}$.
Also, suppose that $\ket{0}_{G} = \sum_{j} \alpha_{j}\ket{v_{j}}$ and $\ket{0}_{H} = \sum_{k} \beta_{k}\ket{w_{k}}$
are the initial states in $G$ and $H$, respectively. Then, the quantum walk on $G \oplus H$ is given by
\begin{equation}
\sum_{j,k} (\alpha_{j}e^{-it\mu_{j}}\ket{v_{j}}) \otimes (\beta_{k}e^{-it\nu_{k}}\ket{w_{k}})
	=
	e^{-itG}\ket{0}_{G} \otimes e^{-itH}\ket{0}_{H}
\end{equation}
This shows that if the mixing times of $G$ and $H$ have a common intersection, then $G \oplus H$ is
instantaneous uniform mixing.
\qed \\

\begin{proposition}
The following graphs are instantaneous uniform mixing:
\begin{enumerate}
\item[(a)] $G^{\oplus k}$, $k \ge 1$, if the weighted graph $G$ is instantaneous uniform mixing.
\item[(b)] Any Cartesian product combinations of $Q_{n}$ and $K_{4}$, for any $n \ge 1$.
\end{enumerate}
\end{proposition}
\prf
(a) Apply Fact \ref{fact:cartesian_product} to $G$ with itself recursively $n-1$ times.
(b) It was shown in \cite{mr02}, the hypercube $Q_{n}$ hits the uniform distribution at times $t = (2k+1)n\pi/4$.
For the complete graphs $K_{n}$, it was proved in \cite{abtw03} that uniform mixing is possible if and only if
\begin{equation}
\frac{4}{n}\sin^{2}\left(\frac{tn}{2(n-1)}\right) = 1.
\end{equation}
So, for $K_{3}$, uniform mixing is achieved if $\sin^{2}(3t/4) = 3/4$ (or $3t/4 = \sin^{-1}(\pm\sqrt{3}/2)$),
and for $K_{4}$, if $\sin^{2}(2t/3) = 1$ (or $t = (2k+1)(3\pi/4)$).
Note that the uniform mixing times of $Q_{n}$ and $K_{4}$ have common intersections.
\qed \\

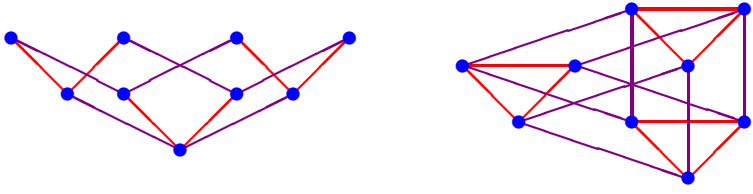
\begin{figure}[t]
\begin{center}
\setlength{\unitlength}{0.75cm}
\begin{picture}(16,4)
\thicklines

\put(2,1.5){\color{red}{\line(-1,1){1}}}
\put(2,1.5){\color{red}{\line(1,1){1}}}
\put(6,1.5){\color{red}{\line(-1,1){1}}}
\put(6,1.5){\color{red}{\line(1,1){1}}}
\put(4,0.5){\color{red}{\line(-1,1){1}}}
\put(4,0.5){\color{red}{\line(1,1){1}}}
\put(4,0.5){\color{purple}{\line(-2,1){2}}}
\put(4,0.5){\color{purple}{\line(2,1){2}}}
\put(3,1.5){\color{purple}{\line(-2,1){2}}}
\put(3,1.5){\color{purple}{\line(2,1){2}}}
\put(5,1.5){\color{purple}{\line(-2,1){2}}}
\put(5,1.5){\color{purple}{\line(2,1){2}}}
\put(1,2.5){\color{blue}{\circle*{0.25}}}
\put(2,1.5){\color{blue}{\circle*{0.25}}}
\put(3,2.5){\color{blue}{\circle*{0.25}}}
\put(3,1.5){\color{blue}{\circle*{0.25}}}
\put(4,0.5){\color{blue}{\circle*{0.25}}}
\put(5,1.5){\color{blue}{\circle*{0.25}}}
\put(5,2.5){\color{blue}{\circle*{0.25}}}
\put(6,1.5){\color{blue}{\circle*{0.25}}}
\put(7,2.5){\color{blue}{\circle*{0.25}}}

\put(10,1){\color{red}{\line(-1,1){1}}}
\put(10,1){\color{red}{\line(1,1){1}}}
\put(9,2){\color{red}{\line(1,0){2}}}
\put(13,0){\color{red}{\line(-1,1){1}}}
\put(13,0){\color{red}{\line(1,1){1}}}
\put(12,1){\color{red}{\line(1,0){2}}}
\put(13,2){\color{red}{\line(-1,1){1}}}
\put(13,2){\color{red}{\line(1,1){1}}}
\put(12,3){\color{red}{\line(1,0){2}}}
\put(9,2){\color{purple}{\line(3,-1){3}}}
\put(9,2){\color{purple}{\line(3,1){3}}}
\put(12,1){\color{purple}{\line(0,1){2}}}
\put(11,2){\color{purple}{\line(3,-1){3}}}
\put(11,2){\color{purple}{\line(3,1){3}}}
\put(14,1){\color{purple}{\line(0,1){2}}}
\put(10,1){\color{purple}{\line(3,-1){3}}}
\put(10,1){\color{purple}{\line(3,1){3}}}
\put(13,0){\color{purple}{\line(0,1){2}}}
\put(9,2){\color{blue}{\circle*{0.25}}}
\put(10,1){\color{blue}{\circle*{0.25}}}
\put(11,2){\color{blue}{\circle*{0.25}}}
\put(12,1){\color{blue}{\circle*{0.25}}}
\put(13,0){\color{blue}{\circle*{0.25}}}
\put(14,1){\color{blue}{\circle*{0.25}}}
\put(12,3){\color{blue}{\circle*{0.25}}}
\put(13,2){\color{blue}{\circle*{0.25}}}
\put(14,3){\color{blue}{\circle*{0.25}}}

\end{picture}
\caption{Examples of instantaneous uniform mixing graphs:
from left to right: (a) $P_{3} \oplus P_{3}$; (b) $K_{3} \oplus K_{3}$.
}
\label{figure:uniform}
\vspace{.1in}
\hrule
\end{center}
\end{figure}

\par\noindent
It is not known if the cycles $C_{n}$, weighted or not, are uniform mixing \cite{abtw03}, 
except for $C_{3} = K_{3}$ and $C_{4} = Q_{2}$. In the following, we show that $C_{5}$ is not uniform mixing.

\begin{fact}
The unweighted $C_{5}$ is not instantaneous uniform mixing.
\end{fact}
\prf
The eigenvalues of $C_{5}$ are $\lambda_{j} = 2\cos(2\pi j/5)$, $j=0,\ldots,4$ (see \cite{biggs}). 
In fact, they exhibit some symmetries since
$\lambda_{0} = 2$, $\lambda_{1} = \lambda_{4} = 2\cos\left(\frac{2\pi}{5}\right) = (-1+\sqrt{5})/2$,
and $\lambda_{2} = \lambda_{3} = 2\cos\left(\frac{4\pi}{5}\right) = (-1-\sqrt{5})/2$.
Let $\lambda_{\pm} = (-1\pm\sqrt{5})/2$; thus, $\lambda_{1} = \lambda_{+}$ and $\lambda_{2} = \lambda_{-}$.

The eigenvectors of $C_{5}$ are $\ket{v_{j}}$, where $\braket{k}{v_{j}} = \omega^{jk}/\sqrt{5}$, for
$j,k = 0,\ldots,4$ and $\omega = \exp(2\pi i/5)$.
Given that $\ket{0} = \frac{1}{\sqrt{5}}\sum_{j=0}^{4} \ket{v_{j}}$, the quantum walk on $C_{5}$ is given by:
\begin{eqnarray}
\ket{\psi(t)} 
	& = & \frac{1}{\sqrt{5}}\left\{e^{-2it}\ket{v_{0}} + e^{-it\lambda_{1}}(\ket{v_{1}} + \ket{v_{4}})
		+ e^{-it\lambda_{2}}(\ket{v_{2}} + \ket{v_{3}})\right\}.
\end{eqnarray}
We note that 
$\ket{v_{1}} + \ket{v_{4}} = \frac{1}{\sqrt{5}}[\lambda_{0} \lambda_{+} \lambda_{-} \lambda_{-} \lambda_{+}]^{T}$
and
$\ket{v_{2}} + \ket{v_{3}} = \frac{1}{\sqrt{5}}[\lambda_{0} \lambda_{-} \lambda_{+} \lambda_{+} \lambda_{-}]^{T}$.
Thus, the amplitude of the quantum walk is given by
\begin{eqnarray}
\braket{0}{\psi(t)} 
	& = & \frac{1}{5} \left\{e^{-it\lambda_{0}} + \sum_{\pm} \lambda_{0}e^{-it\lambda_{\pm}} \right\} \\
\braket{1}{\psi(t)} = \braket{4}{\psi(t)}
	& = & \frac{1}{5} \left\{e^{-it\lambda_{0}} + \sum_{\pm} \lambda_{\pm}e^{-it\lambda_{\pm}} \right\} \\
\braket{2}{\psi(t)} = \braket{3}{\psi(t)}
	& = & \frac{1}{5} \left\{e^{-it\lambda_{0}} + \sum_{\pm} \lambda_{\mp}e^{-it\lambda_{\pm}} \right\}
\end{eqnarray}
Let $\mu_{\pm} = (5 \pm \sqrt{5})/2$. After simplifications, the probability function is given by:
\begin{eqnarray}
p_{0}(t) & = & \frac{1}{25} \left\{9 + 4\sum_{\pm} \cos(\mu_{\pm}t) + 8\cos(\sqrt{5}t)\right\} \\
p_{1}(t) = p_{4}(t) 
	& = & \frac{1}{25} \left\{4 + \sum_{\pm} 2\lambda_{\mp}\cos(\mu_{\pm}t) - 2\cos(\sqrt{5}t)\right\} \\
p_{2}(t) = p_{3}(t) 
	& = & \frac{1}{25} \left\{4 + \sum_{\pm} 2\lambda_{\pm}\cos(\mu_{\pm}t) - 2\cos(\sqrt{5}t)\right\}
\end{eqnarray}
Assume that $C_{5}$ has instantaneous uniform mixing at time $T$. From $p_{1}(T) = p_{2}(T)$, we get
$\cos(\mu_{+}T) = \cos(\mu_{-}T)$ which implies that $\mu_{+} = 2\pi m \pm \mu_{-}$, for some $m \in \mathbb{Z}$.
So, we get either $\sqrt{5}T = 2\pi m$ or $5T = 2\pi m$. 
From $p_{0}(T) = p_{2}(T)$, we get
\begin{equation}
5 + (5 + \sqrt{5})\cos(\mu_{-}T) + (5 - \sqrt{5})\cos(\mu_{+}T) + 10\cos(\sqrt{5}T) = 0.
\end{equation}
If $\sqrt{5}T = 2\pi m$, then 
$(5 + \sqrt{5})\cos(\mu_{-}T) + (5 - \sqrt{5})\cos(\mu_{+}T) + 15 = 0$,
which is a contradiction.
On the other hand, if $5T = 2\pi m$, 
we have $5 \pm 10\alpha + 10(2\alpha^{2} - 1) = 0$, by letting $\alpha = \cos(\sqrt{5}T/2)$.
This implies that $\alpha = (\mp 1 \pm \sqrt{5})/4$ which equals to $\cos(\pi j/5)$, 
for some $j \in \mathbb{Z}^{+}$. Since $p_{0}(T) = p_{2}(T)$, we get $\cos(\sqrt{5}T/2) = \cos(\pi j/5)$;
thus, $\sqrt{5}T/2 = \pi n/5$, for some $n \in \mathbb{Z}$. Also, since $5T = 2\pi m$,
we have $\sqrt{5}T/2 = \pi m/\sqrt{5}$ or $5m/n = \sqrt{5}$, which is a contradiction.
\qed


\section{Average Mixing}

In this section we prove that no weighted graphs are average universal mixing and show
a necessary condition for a weighted graph to be average uniform mixing. But, first we prove a lemma
on the average probability of the start vertex in a quantum walk on any weighted graph.

\begin{lemma}
In a  quantum walk on a weighted graph $G=(V,E)$ starting at an arbitrary vertex, 
the average probability of the start vertex satisfies:
\begin{equation}
\overline{p}_{\mathsf{start}} \geq \frac{1}{\tau(G)}.
\end{equation}
\end{lemma}
\prf
Since the adjacency matrix $A$ of $G$ is a real symmetric matrix, it has real eigenvalues and is real 
orthogonally diagonalizable (see \cite{hj}). Let $\lambda_{k}$ and $\ket{v_{k}}$ be the eigenvalues and
orthonormal eigenvectors of $A$, $k=1,\ldots,n$. Assuming that the start vertex is $0$, without loss of
generality, and that $\ket{0} = \sum_{k} \alpha_{k} \ket{v_{k}}$, for $\alpha_{k} \in \mathbb{R}$, we have
$\sum_{k} \alpha_{k}^{2} = 1$. In what follows, let $\beta_{k} = \alpha_{k}^{2}$.
The quantum walk on $G$ is given by $\ket{\psi(t)} = \sum_{k} e^{-it\lambda_{k}}\alpha_{k} \ket{v_{k}}$.
Thus, the amplitude of the start vertex at time $t$ is $\psi_{0}(t) = \sum_{k} e^{-it\lambda_{k}}\beta_{k}$;
and, the average probability of the start vertex is
\begin{eqnarray}
\overline{p}_{0} 
	& = & \lim_{T \rightarrow \infty} \frac{1}{T}\int_{0}^{T} \dt \
		\sum_{j,k} e^{-it(\lambda_{j}-\lambda_{k})} \beta_{j}\beta_{k} 
	= \sum_{j,k} \iverson{\lambda_{j}=\lambda_{k}} \beta_{j}\beta_{k} \\
	& = & \sum_{\lambda \in Sp(G)} \sum_{j,k} \iverson{\lambda_{j} = \lambda_{k} = \lambda} \beta_{j}\beta_{k} 
	= \sum_{\lambda \in Sp(G)} B_{\lambda}^{2},
\end{eqnarray}
where $B_{\lambda} = \sum_{j:\lambda_{j}=\lambda} \beta_{j}$.
Since $\sum_{\lambda} B_{\lambda} = 1$, the last expression is minimized when $B_{\lambda} = 1/\tau(G)$,
for each $\lambda \in Sp(G)$. Thus, the average probability of the start vertex is at least $1/\tau(G)$.
\qed \\

\par\noindent
The previous lemma has two direct implications to uniform and universal mixings.
In \cite{lrsstw06}, it was proved that if a circulant graph $G$ has bounded eigenvalue multiplicity then
$G$ is average almost-uniform mixing. The next claim shows a partial converse to this for arbitrary weighted
graphs, and thus provides a nearly tight characterization of circulant graphs that are average almost-uniform
mixing. This is because if a graph has bounded eigenvalue multiplicity then it has a linear spectral type;
but the converse if not known to hold, even for the case of circulant graphs.

\begin{corollary}
If a weighted graph $G=(V,E)$ is average almost-uniform mixing then $\tau(G) = O(n)$.
\end{corollary}
\prf
If $\tau(G) = o(n)$, then the average probability of the start vertex is $\omega(1/n)$, which implies
that $G$ is not average almost-uniform mixing.
\qed \\

\begin{corollary}
No weighted graphs are average universal mixing.
\end{corollary}
\prf
Since the average probability of the start vertex is at least $1/\tau(G)$, it is bounded away from zero.
\qed


\section{Conclusions}

In this work, we investigate the set of probability distributions generated by a continuous-time quantum walk on
weighted graphs. We show that the instantaneous probability distributions generated by a quantum walk on the 
weighted claw (or star) graph $K_{1,n}$ ranges over all distributions as the edge weights are varied over the 
non-negative real numbers. In this sense, the weighted claw has the {\em universal} mixing property. 
This is a generalization of the {\em uniform} mixing property on unweighted graphs considered in earlier works 
on the hypercube \cite{mr02}, the complete graphs \cite{abtw03}, and the Cayley graph of the symmetric group 
\cite{gw03}. Our next result shows that all complete multipartite graphs are universal mixing. This stands
in contrast with the fact that unweighted complete multipartite graphs are not uniform mixing, except for the
lone case of $K_{2,2}$ (see \cite{abtw03}). The proof of the multipartite result uses a weighted generalization
of the {\em path collapsing} argument (from \cite{ccdfgs03}).
These results on instantaneous universal mixing of weighted graphs can be extended to unweighted 
multigraphs (where multiple edges can connect two vertices) if an approximate mixing notion is allowed. 

For universal mixing over average distributions, we show that there are no graphs with this property.
In fact, a key ingredient in this proof shows a necessary condition for a graph to be average almost-uniform
mixing. A weighted graph is average almost-uniform mixing unless its spectral type is sublinear in the
number of vertices. This provides a near tight characterization for circulant graphs, since they are known 
to be average almost-uniform mixing if the eigenvalues have bounded multiplicities \cite{lrsstw06}. 
Note that bounded eigenvalue multiplicities implies linear spectral type; but the converse is unclear, 
even for circulants.

A main open question left from this work is whether weighted paths $P_{n}$, $n \ge 4$, are instantaneous 
universal mixing. If the weighted paths $P_{n}$ are universal mixing, then so are all weighted trees; 
but if they are not, then an interesting question is to characterize the weighted trees that are universal mixing. 
A related question on weighted paths is whether they are average almost-uniform mixing, given that their 
spectral type is always linear (see \cite{jl99}). We leave these questions for future work.


\section*{Acknowledgments}

This research was supported by the National Science Foundation grant DMS-0353050 while the authors were
part of the Clarkson-Potsdam Research Experience for Undergraduates (REU) Summer program in Mathematics
at State University of New York at Postdam. The research of C. Tamon was also supported by the
National Science Foundation grant DMR-0121146 through the Center for Quantum Device Technology at 
Clarkson University.


\end{document}

\newpage

\appendix

\section{The analysis of $P_{n}$}

The adjacency matrix of a weighted $P_{4}$ is given by:
\begin{equation}
A = 	\begin{bmatrix}
	0 & 1 & 0 & 0 \\
	1 & 0 & a & 0 \\
	0 & a & 0 & b \\
	0 & 0 & b & 0
	\end{bmatrix}
\end{equation}
Define the following three expressions:
\begin{eqnarray}
\alpha_{1} & = & \sqrt{b^{4} - 2b^{2} + 1 + a^{4} + 2a^{2}b^{2} + 2a^{2}} \\
\alpha_{2} & = & 2a^{2} + 2b^{2} + 2 - 2\alpha_{1} \\
\alpha_{3} & = & 2a^{2} + 2b^{2} + 2 + 2\alpha_{1}
\end{eqnarray}
The eigenvalues of $A$ are given by:
\begin{equation}
\mu_{\pm} = \pm\frac{1}{2}\sqrt{\alpha_{3}}, \ \ \
\nu_{\pm} = \pm\frac{1}{2}\sqrt{\alpha_{2}}
\end{equation}
with eigenvectors given by:
\begin{equation}
\ket{v_{\pm}} =  
	\begin{bmatrix} 
	\frac{\pm 1}{8ab}\sqrt{\alpha_{3}}(4-\alpha_{2}) \\
	\frac{1}{4ab}(\alpha_{3}-4b^{2}) \\
	\frac{\pm 1}{2b}\sqrt{\alpha_{3}} \\
	1
	\end{bmatrix}, \ \ \
\ket{w_{\pm}} =
	\begin{bmatrix} 
	\frac{\pm 1}{8ab}\sqrt{\alpha_{2}}(4-\alpha_{3}) \\ 
	\frac{1}{4ab}(\alpha_{2}-4b^{2}) \\
	\frac{\pm 1}{2b}\sqrt{\alpha_{2}} \\
	1
	\end{bmatrix}
\end{equation}
Let $M = \sqrt{\braket{v_{\pm}}{v_{\pm}}}$ and $N = \sqrt{\braket{w_{\pm}}{w_{\pm}}}$.
Since
$\ket{0} = \sum_{\pm} \braket{0}{v_{\pm}}\ket{v_{\pm}} + \sum_{\pm} \braket{0}{w_{\pm}}\ket{w_{\pm}}$,
in a quantum walk on $P_{3}$ starting at the leftmost vertex $0$, 
the amplitude of the rightmost vertex $3$, is given by:
\begin{eqnarray}
\bra{3}e^{-itA}\ket{0} 
	& = &
	\sum_{\pm} e^{-it\mu_{\pm}} \braket{0}{v_{\pm}} \braket{3}{v_{\pm}} 
	+ 
	\sum_{\pm} e^{-it\nu_{\pm}} \braket{0}{w_{\pm}} \braket{3}{w_{\pm}} \\
	& = &
	\sum_{\pm} e^{-it\mu_{\pm}} (\pm \braket{0}{v_{+}}) \braket{3}{v_{\pm}} 
	+ 
	\sum_{\pm} e^{-it\nu_{\pm}} (\pm \braket{0}{w_{+}}) \braket{3}{w_{\pm}}, \\
	& & 
	\mbox{ since $\braket{0}{v_{\pm}} = \pm\braket{0}{v_{+}}$ and similarly for $\ket{w_{\pm}}$ } \\
	& = & -2i
	\left\{\braket{0}{v_{+}}\braket{3}{v_{+}}\sin(\mu t)
	+
	\braket{0}{w_{+}}\braket{3}{w_{+}}\sin(\nu t)\right\}, \\
	& & 
	\mbox{ since $\braket{3}{w_{+}} = \braket{3}{w_{-}}$ }.
\end{eqnarray}
This shows that the amplitude of the last vertex is {\em always} imaginary, since the
other expressions are all real.

\end{document}